\documentclass[preprint,showpacs,nofootinbib,prd,aps]{revtex4-1}
\usepackage{graphicx,amstext,amssymb}

\begin{document}
\title{Quantum canonical ensemble  and correlation femtoscopy at fixed multiplicities}

\author{S.V. Akkelin$^{1}$}
\author{Yu.M. Sinyukov$^{1,2}$}
\affiliation{$^1$Bogolyubov Institute for Theoretical Physics,
Metrolohichna  14b, 03680 Kiev,  Ukraine\\
 $^2$ExtreMe Matter Institute EMMI, GSI~Helmholtzzentrum f\"ur~Schwerionenforschung,
D-64291 Darmstadt, Germany }

\begin{abstract}

Identical particle correlations at fixed multiplicity are considered
by means of quantum canonical ensemble of finite systems. We
calculate one-particle momentum spectra and two-particle
Bose-Einstein correlation functions  in the ideal gas by using a
recurrence relation for the partition function. Within such a model
we investigate the validity of the thermal Wick's theorem  and its
applicability for decomposition of the two-particle distribution
function. The dependence of the Bose-Einstein correlation parameters
on the average momentum of the particle pair is also investigated.
Specifically, we present the analytical formulas that allow one to
estimate the effect of suppressing the correlation functions in a
finite canonical system. The results can be used for the femtoscopy
analysis of the $A+A$ and $p+p$ collisions with selected (fixed)
multiplicity.

\end{abstract}


 \maketitle

 \section{Introduction}

The correlation femtoscopy  method (for reviews see, e.g., Refs.
\cite{KopPodg,HBT,Sin-1})) uses  momentum correlations of identical
particles to extract information about the spatiotemporal structure
of extremely small and short-lived systems created in nucleon  and
nuclear collisions. The method  is grounded in the Bose-Einstein or
Fermi-Dirac symmetric properties of the quantum states. Because  in
high-energy nucleus-nucleus or hadron-hadron collisions most of
produced particles are pions, the  Bose-Einstein correlations of two
identical pions are usually analyzed to increase the statistics of
the correlation femtoscopy measurements. Inasmuch as mean particle
multiplicities increase with collision energy, one can divide a
whole set of high energy collision events into subsets with  fixed
charged-particle multiplicities. In recent papers  \cite{Heinz} it
was considered even a possibility of  single-event correlation
femtoscopy -- at least, theoretically, because in reality  not
enough pairs remain for a statistically meaningful analysis in a
single event. Typically, the ensemble  of events with
charged-particle numbers selected in some fixed multiplicity bins is
analyzed. In heavy-ion collisions, the measurement of observables as
a function of the multiplicity class  has a long history and is
regarded as a proxy for centrality dependence. Recently, due to the
start of  Large Hadron Collider (LHC) experiments where colliding
energies per nucleon pair are in a TeV region, the fixed particle
multiplicity technique has also been utilized  for analysis of the
Bose-Einstein correlations of identical particles in proton-proton
collisions \cite{HBT-pp}.

It is firmly  established now that high-energy heavy-ion collisions
are  described basically by relativistic hydrodynamics (for recent
reviews, see Ref. \cite{Hydro}), and that at the later dilute stage
of matter evolution the hydrodynamics is followed by the highly
dissipative hadronic gas expansion that is modeled by a hadronic
cascade model like UrQMD \cite{urqmd} till the free streaming regime
is reached. The so-called particlization -- transition from
continuous medium (hydrodynamic) consideration of the system to its
particle-based description -- is associated typically with the
lowest possible temperature when the system is still close to local
thermal and chemical equilibrium. It defines the isotherm that  is
often called the chemical freeze-out hypersurface. The situation is
less clear in high-multiplicity proton-proton collisions, and there
are also attempts to describe particle momentum spectra in such
collisions using a hydrodynamic approach (see, e.g., Ref.
\cite{flow-pp}).

It is worth to noting  that quantum statistics is not an inherent
property of these, as well as  many other,  quasiclassical models.
Typically, when hadrons are generated on the particlization
hypersurface, the single particle weight is sampled according to the
Bose-Einstein or Fermi-Dirac distributions in the grand canonical
ensemble (with viscous corrections, if necessary) in the framework
of the so-called Cooper-Frye prescription \cite{CFp}.  As for the
two-identical-particle  spectra, the current quasiclassical
simulations utilize the factorized decomposition of the two-particle
emission function into the single-particle ones with the additional
multiplier proportional to  the module squared of two-particle
symmetrized or antisymmetrized amplitudes, so that the correlation
function becomes (omitting all nonprincipal details) $C(p_1,
p_2)\propto\left\langle 1\pm\cos(p_1-p_2)(x_1-x_2)\right\rangle$,
where angular brackets mean averaging  over an emission function.
Such a local ``switching on'' of the quantum statistic effects in
two-particle cross sections is just like  in the
final-state-interaction  method.  However, the quantum statistics is
not associated with local  two-particle interaction but is the
global effect, and complete symmetrization or antisymmetrization of
the {\it total } system is required to find the correct results for
one- and multi- particle momentum spectra. Such an analysis was done
in Refs. \cite{Sin,Ledn}. It has been shown that for thermal
identical particles the above mentioned procedure for one- and
two-boson spectra evaluations in quasiclassical models is correct if
one provides the symmetrization in an {\it ensemble} of initially
independently emitted thermal Boltzmann particles with the Poisson
distribution for the particle numbers in the ensemble. Otherwise,
the above-described prescription for the simulation of the single-
and double-particle spectra and the  correlation function
$C(p_1,p_2)$ is not correct. In particular, it is violated  for the
states with a fixed boson number. The detailed grounding of such a
conclusion, provided in Ref. \cite{Ledn} with corresponding
numerical calculations, is based on the nonrelativistic
Kopylov-Podgoretsky model \cite{KopPodg} of initially Boltzmann
independent factorized sources (with subsequent symmetrization).

In this article  we use as a basis  thermal canonical and
grand-canonical bosonic ensembles for quantum relativistic ideal
gases, which  allows us to avoid  the specific procedure of
``switching on'' the quantum statistics as well as the assumption of
initially distinguishable sources. Our aim is to clarify the general
reasons of violation of the standard prescriptions for $C(p_1,p_2)$
(see above) when one calculates the correlation function in events
with fixed multiplicity or just on an event-by-event basis
\cite{Heinz}. The answer is not trivial and depends on the
applicability of the thermal Wick's theorem \cite{Wick-1}. It is
beyond the scope of this study to give a comprehensive analysis and
prescription for quasiclassical event generators dealing with $p+p$
and $A+A$ collisions. Rather, the purpose of this paper is to show
how imposed particle number constraints affect the single-particle
momentum density and the two-particle momentum correlation function
in a canonical ensemble of a finite system, as well as to present
the analytical estimates for these values in some tractable
approximations. In particular,  the conditions under which the
standard decomposition of two-particle distribution can be applied
are considered.

\section{One- and two- particle  momentum spectra in grand-canonical and canonical ensembles of identical bosons}

We begin with a brief overview of  the properties of the quantum
grand-canonical ensemble of a noninteracting  boson field  with
plane waves satisfying  periodic boundary conditions on the walls of
a cubic box (see, e.g., Ref. \cite{Bog}).

\subsection{Momentum spectra and Wick's theorem in a grand-canonical ensemble}

The basic object, a grand-canonical statistical operator, can be
written as follows:
\begin{eqnarray}
\rho=\exp{(-\beta (\widehat{H}-\mu \widehat{N})}),  \label{1}
\end{eqnarray}
where  $\beta = 1/T$ is the inverse temperature,
$\widehat{H}=\sum_{p}\epsilon_{p}a^{+}_{p}a_{p}$ is the Hamiltonian,
$\epsilon_{p}$ is the energy of the single-particle state, and
$\widehat{N}=\sum_{p}a^{+}_{p}a_{p}$ is the particle number
operator. Creation, $a^{+}_{p}$, and annihilation, $a_{p}$,
operators satisfy the following canonical commutation relation in
the discrete-mode representation:
\begin{eqnarray}
[a_{p}, a^{+}_{p'}] = \delta_{pp'}, \label{2}
\end{eqnarray}
where $\delta_{pp'}$ is the Kronecker delta function. For notational
simplicity, here and below we write $p$ instead of
$(p_{x},p_{y},p_{z})$. The expectation value of the  operator
$\widehat{A}$ can be expressed as
\begin{eqnarray}
\langle  \widehat{A }\rangle = \frac{Tr[\rho  \widehat{A}]}{Z },
\label{3}
\end{eqnarray}
where $Z $ is the grand-canonical partition function,
\begin{eqnarray}
Z =Tr[\rho]. \label{3.1}
\end{eqnarray}
In what follows  we assume for simplicity that the chemical
potential $\mu = 0$. For $\mu \neq 0$ the  substitution
$\epsilon_{p} \rightarrow \epsilon_{p} - \mu$ has to be utilized in
corresponding expressions.

Using the eigenstates of the particle number operator,
\begin{eqnarray}
|p_{1},...,p_{N}\rangle =
\frac{1}{\sqrt{N!}}a^{+}_{p_{1}}...a^{+}_{p_{N}}|0\rangle ,
\label{4}
\end{eqnarray}
and the identity
\begin{eqnarray}
\sum_{N}\sum_{p_{1},...,p_{N}}|p_{1},...,p_{N}\rangle \langle
p_{1},...,p_{N} |= 1, \label{5}
\end{eqnarray}
one can write the statistical operator (\ref{1}) in the form
\begin{eqnarray}
\rho =  \sum_{N}\sum_{p_{1},...,p_{N}}e^{-\beta \epsilon_{p_1} - ...
- \beta \epsilon_{p_{N}}} |p_{1},...,p_{N}\rangle \langle
p_{1},...,p_{N} | . \label{6}
\end{eqnarray}
Inasmuch as our aim here is to calculate the two-boson correlation
function, we are interested in the expectation values of operators
$a^{+}_{p_{1}}a_{p_{2}}$ and
$a^{+}_{p_{1}}a^{+}_{p_{2}}a_{p_{1}}a_{p_{2}}$; other $n-$point
operator functions can be calculated in a similar way, if necessary.
For calculations we adapt  the method proposed in Ref. \cite{Wick-1}
(see also Refs. \cite{Bog,Groot}). The corresponding results are, of
course, well known, but it allows us to show a  simple example of
calculations within the method of Ref. \cite{Wick-1} and will help
to reveal  differences between calculations with and without a fixed
particle number constraint.

Our starting point is the relationship
\begin{eqnarray}
a_{p} \rho = \rho a_{p} e^{-\beta \epsilon_{p}}, \label{10}
\end{eqnarray}
which can be proved  by using an elementary operator algebra and Eq.
(\ref{6}). Then, using trace invariance under the cyclic permutation
of an operator, we get
\begin{eqnarray}
Tr[\rho a^{+}_{p_{1}}a_{p_{2}}] = e^{-\beta \epsilon_{p_{2}}}
Tr[\rho a_{p_{2}} a^{+}_{p_{1}}]= e^{-\beta \epsilon_{p_{2}}}Tr[\rho
a^{+}_{p_{1}}a_{p_{2}}]+ e^{-\beta \epsilon_{p_{2}}}\delta_{p_1p_2}
Tr[\rho ]. \label{11}
\end{eqnarray}
From the above equation we have
\begin{eqnarray}
\langle a^{+}_{p_{1}}a_{p_{2}}\rangle  =
\frac{\delta_{p_1p_2}}{e^{\beta \epsilon_{p_{2}}}-1 }, \label{12}
\end{eqnarray}
which  is a familiar Bose-Einstein distribution. Similarly, the
trace $Tr[\rho a^{+}_{p_{1}}a^{+}_{p_{2}}a_{p_{1}}a_{p_{2}}]$ can be
expressed as
\begin{eqnarray}
Tr[\rho a^{+}_{p_{1}}a^{+}_{p_{2}}a_{p_{1}}a_{p_{2}}] =  \nonumber \\
 e^{-\beta
\epsilon_{p_{2}}}\delta_{p_1p_2} Tr[\rho a^{+}_{p_{2}}a_{p_{1}}]+
e^{-\beta \epsilon_{p_{2}}}\delta_{p_2p_2}Tr[\rho
a^{+}_{p_{1}}a_{p_{1}}] + e^{-\beta \epsilon_{p_{2}}}Tr[\rho
a^{+}_{p_{1}}a^{+}_{p_{2}}a_{p_{1}}a_{p_{2}}], \label{13}
\end{eqnarray}
and we have
\begin{eqnarray}
Tr[\rho a^{+}_{p_{1}}a^{+}_{p_{2}}a_{p_{1}}a_{p_{2}}] =
\frac{\delta_{p_1p_2}}{e^{\beta \epsilon_{p_{2}}}-1 } Tr[\rho
a^{+}_{p_{2}}a_{p_{1}}]+ \frac{\delta_{p_2p_2}}{e^{\beta
\epsilon_{p_{2}}}-1 }Tr[\rho a^{+}_{p_{1}}a_{p_{1}}].  \label{14}
\end{eqnarray}
Then, taking into account Eq. (\ref{12}), $\langle
a^{+}_{p_{1}}a^{+}_{p_{2}}a_{p_{1}}a_{p_{2}}\rangle $ reads
\begin{eqnarray}
\langle a^{+}_{p_{1}}a^{+}_{p_{2}}a_{p_{1}}a_{p_{2}}\rangle =
\langle a^{+}_{p_{2}}a_{p_{1}}\rangle \langle
a^{+}_{p_{1}}a_{p_{2}}\rangle + \langle
a^{+}_{p_{1}}a_{p_{1}}\rangle \langle a^{+}_{p_{2}}a_{p_{2}}\rangle
, \label{15}
\end{eqnarray}
which is nothing but the particular case of the  thermal Wick's
theorem \cite{Wick-1} (see also Ref. \cite{Wick-2} where the
applicability of the thermal Wick's theorem for inhomogeneous
locally equilibrated noninteracting systems is analyzed). Note  that
to derive Eqs. (\ref{12}) and (\ref{15}) we do not need an explicit
expression for the grand-canonical partition function (\ref{3.1}).

\subsection{The momentum spectra in a canonical ensemble with a fixed particle number constraint (discrete-mode representation) }

Now, let us apply the fixed particle number constraint to the
grand-canonical statistical operator (\ref{1}). For this aim, one
can utilize  the projection operator ${\cal P}_{N}$,
\begin{eqnarray}
{\cal P}_{N} = \sum_{p_{1},...,p_{N}} |p_{1},...,p_{N}\rangle
\langle p_{1},...,p_{N} | ,  \label{7}
\end{eqnarray}
which  automatically invokes the corresponding constraint. Then we
assert that the canonical statistical operator with the constraint,
$\rho_{N}$, is
\begin{eqnarray}
\rho_{N} = {\cal P}_{N}\rho {\cal P}_{N}
=\sum_{p_{1},...,p_{N}}e^{-\beta \epsilon_{p_1} - ... - \beta
\epsilon_{p_{N}}} |p_{1},...,p_{N}\rangle \langle p_{1},...,p_{N} |
. \label{8}
\end{eqnarray}
The expectation value of the operator $\widehat{A}$ can be defined
as
\begin{eqnarray}
\langle  \widehat{A }\rangle_{N} = \frac{Tr[\rho_{N}  A]}{Z_{N} },
\label{9}
\end{eqnarray}
where $Z_{N} $ is the corresponding  canonical partition function,
\begin{eqnarray}
Z_{N} =Tr[\rho_{N}]. \label{9.1}
\end{eqnarray}
To evaluate  the expectation values of operators
$a^{+}_{p_{1}}a_{p_{2}}$ and
$a^{+}_{p_{1}}a^{+}_{p_{2}}a_{p_{1}}a_{p_{2}}$ with the canonical
statistical operator $\rho_{N}$, see Eqs. (\ref{8}) and (\ref{9}),
 we first utilize elementary operator algebra
to prove that
\begin{eqnarray}
a_{p} \rho_{N} = \rho_{N-1} a_{p} e^{-\beta \epsilon_{p}}.
\label{16}
\end{eqnarray}
Then, to evaluate the trace $Tr[\rho_{N} a^{+}_{q_{1}}a_{q_{2}}]$,
we exploit its invariance under cyclic permutations and  get
\begin{eqnarray}
Tr[\rho_{N} a^{+}_{p_{1}}a_{p_{2}}] = e^{-\beta \epsilon_{p_{2}}}
Tr[\rho_{N-1} a_{p_{2}} a^{+}_{p_{1}}]= e^{-\beta
\epsilon_{p_{2}}}Tr[\rho_{N-1} a^{+}_{p_{1}}a_{p_{2}}]+ e^{-\beta
\epsilon_{p_{2}}}\delta_{p_1p_2}Tr[\rho_{N-1} ]. \label{17}
\end{eqnarray}
From Eq. (\ref{17}) we have the iteration relation
\begin{eqnarray}
\langle a^{+}_{p_{1}}a_{p_{2}}\rangle_{N}  = e^{-\beta
\epsilon_{p_{2}}}\delta_{p_1p_2}\frac{Z_{N-1}}{Z_{N}}+ e^{-\beta
\epsilon_{p_{2}}}\frac{Z_{N-1}}{Z_{N}}\langle
a^{+}_{p_{1}}a_{p_{2}}\rangle_{N-1}. \label{18}
\end{eqnarray}
By using Eq. (\ref{18}) one  can prove by induction (see also Ref.
\cite{Recurr-2} and references therein) that
\begin{eqnarray}
\langle a^{+}_{p_{1}}a_{p_{2}}\rangle_{N}  =
\delta_{p_1p_2}\sum_{i=1}^{N}  e^{-i\beta
\epsilon_{p_{2}}}\frac{Z_{N-i}}{Z_{N}}.  \label{19}
\end{eqnarray}
In the same way  we obtain
\begin{eqnarray}
\langle  a^{+}_{p_{1}}a^{+}_{p_{2}}a_{p_{1}}a_{p_{2}}\rangle_{N}  =
e^{-\beta \epsilon_{p_{2}}}\frac{Z_{N-1}}{Z_{N}}\langle a_{p_{2}}
a^{+}_{p_{1}}a^{+}_{p_{2}}a_{p_{1}}\rangle_{N-1} = \nonumber
\\ e^{-\beta \epsilon_{p_{2}}}\frac{Z_{N-1}}{Z_{N}}\left(\langle
a^{+}_{p_{1}}a^{+}_{p_{2}}a_{p_{1}}a_{p_{2}}\rangle_{N-1} +
\delta_{p_1p_2}\langle a^{+}_{p_{2}}a_{p_{1}}\rangle_{N-1}+
\delta_{p_2p_2}\langle a^{+}_{p_{1}}a_{p_{1}}\rangle_{N-1}\right  ),
\label{20}
\end{eqnarray}
and one can prove by induction that
\begin{eqnarray}
\langle  a^{+}_{p_{1}}a^{+}_{p_{2}}a_{p_{1}}a_{p_{2}}\rangle_{N}  =
\delta_{p_1p_2}\sum_{i=1}^{N}  e^{-i\beta
\epsilon_{p_{2}}}\frac{Z_{N-i}}{Z_{N}}\langle
a^{+}_{p_{2}}a_{p_{1}}\rangle_{N-i}+ \delta_{p_2p_2}\sum_{i=1}^{N}
e^{-i\beta \epsilon_{p_{2}}}\frac{Z_{N-i}}{Z_{N}}\langle
a^{+}_{p_{1}}a_{p_{1}}\rangle_{N-i}. \label{21}
\end{eqnarray}
Finally, using  Eq. (\ref{19}) and  taking into account that
$\langle a^{+}_{p_{i}}a_{p_{j}}\rangle_{0}=0$, we see that Eq.
(\ref{21}) becomes
\begin{eqnarray}
\langle  a^{+}_{p_{1}}a^{+}_{p_{2}}a_{p_{1}}a_{p_{2}}\rangle_{N}  =
(\delta_{p_1p_2}\delta_{p_2p_1} +
\delta_{p_2p_2}\delta_{p_1p_1})\sum_{i=1}^{N-1} \sum_{j=1}^{N-i}
e^{-i\beta \epsilon_{p_{2}}}e^{-j \beta
\epsilon_{p_{1}}}\frac{Z_{N-i-j}}{Z_{N}}. \label{21.1}
\end{eqnarray}

It is immediately apparent from Eqs. (\ref{19}), (\ref{21}), and
(\ref{21.1}) that for the noninteracting canonical ensemble with the
fixed particle number constraint the decomposition (\ref{15}), which
follows from the thermal Wick's theorem, is no more
valid.\footnote{Note  also Ref. \cite{Recurr-3}, where expressions
for these expectation values are derived  in rather compact form at
the price of an additional integration.} Also, notice that for
practical utilizations of  Eqs. (\ref{19}) and (\ref{21.1}) one
needs first to calculate the canonical partition functions. The
latter can be done by means of the recurrence relations as given in
Ref. \cite{Recurr-1}. Below, for the reader's convenience, we
present an elementary derivation of it. As the starting point we
utilize the  relation
\begin{eqnarray}
\sum_{p}\langle  a^{+}_{p}a_{p}\rangle_{N}  = N, \label{22}
\end{eqnarray}
which follows from the definition of $\rho_N$, see Eqs. (\ref{8})
and (\ref{9}). Then, accounting for Eq. (\ref{19}) we get
\begin{eqnarray}
N Z_{N}= \sum_{i=1}^{N} \sum_{p} e^{-i\beta \epsilon_{p}}Z_{N-i}.
\label{23}
\end{eqnarray}
The above expression can be rewritten as
\begin{eqnarray}
Z_{N}= \frac{1}{N}\sum_{i=0}^{N-1}Z_{i} \sum_{p} e^{-(N-i)\beta
\epsilon_{p}}. \label{24}
\end{eqnarray}
Taking into account that $Z_{0}=\langle 0 | 0 \rangle =1$, we get
$Z_{1}=\sum_{p} e^{- \beta \epsilon_{p}}$. Then the recurrence
relation can be expressed in its final form  as
\begin{eqnarray}
Z_{N} = \frac{1}{N}\sum_{i=0}^{N-1}Z_{i} Z_{1}((N-i)\beta),
\label{25}
\end{eqnarray}
where for notational convenience we have defined the quantities
\begin{eqnarray}
Z_{1}(j\beta)=\sum_{p} e^{- j\beta \epsilon_{p}}. \label{25.1}
\end{eqnarray}
Then, the use of Eqs. (\ref{25}) and (\ref{25.1}) allows one to
determine the canonical partition functions and, therefore, to
calculate Eqs. (\ref{19}) and (\ref{21.1}) in the canonical ensemble
with periodical boundary conditions.

\subsection{Canonical ensemble in the thermodynamic limit}

It is worth noting that  periodical boundary conditions are just a
mathematical trick that allows convenient mathematical description
of finite-volume systems. To eliminate these artificial assumptions,
the transition to the thermodynamic limit is typically applied.
Then, the number (or the mean number in a grand-canonical ensemble)
of particles $N$ and the volume $V$ of the system go to $\infty$
while keeping the particle density $n=N/V$ constant. As a  result,
the discrete-mode representation tends to the continuous momentum
representation of the canonical operators.  In this limit, strictly
speaking, normalization of the statistical operator fails because
the partition function diverges, but the expectation values can
still be defined. There are no problems with the application of this
limit to the expectation values (\ref{12}) and (\ref{15}) calculated
in the grand-canonical ensemble, but the computation of (\ref{19})
and (\ref{21})  in the canonical ensemble is  a more involved
problem because Eqs. (\ref{19}) and (\ref{21}) explicitly depend on
canonical partition functions. To overcome this problem, let us
first utilize Eq. (\ref{23})  to write $Z_{N-j-1}/Z_{N-j}$ as
\begin{eqnarray}
\frac{Z_{N-j-1}}{Z_{N-j}} =
\frac{N-j}{N-j-1}\frac{\sum_{i=1}^{N-j-1} \sum_{p} e^{-i\beta
\epsilon_{p}}Z_{N-j-i-1}} { \sum_{i=1}^{N-j} \sum_{p} e^{-i\beta
\epsilon_{p}}Z_{N-j-i}}. \label{34-0}
\end{eqnarray}
It is immediately apparent from Eq. (\ref{34-0}) that for any fixed
$j$ in the thermodynamic limit
\begin{eqnarray}
\lim_{N,V \rightarrow \infty}\frac{Z_{N-j-1}}{Z_{N-j}}=\lim_{N,V
\rightarrow \infty}\frac{Z_{N-1}}{Z_{N}}\equiv \gamma,
\label{34-0-1}
\end{eqnarray}
and therefore,
\begin{eqnarray}
\lim_{N,V \rightarrow \infty}\frac{Z_{N-j}}{Z_{N}}=\lim_{N,V
\rightarrow \infty}\left ( \frac{Z_{N-1}}{Z_{N}} \right )^{j}.
\label{34-00}
\end{eqnarray}
Thus we observe that  Eq. (\ref{34-0}) in the thermodynamic limit
becomes the identity,
\begin{eqnarray}
\frac{Z_{N-1}}{Z_{N}} = \frac{\sum_{i=1}^{\infty} \sum_{p}
e^{-i\beta \epsilon_{p}}\left (\frac{Z_{N-1}}{Z_{N}}\right )^{i+1}}
{ \sum_{i=1}^{\infty} \sum_{p} e^{-i\beta \epsilon_{p}}\left
(\frac{Z_{N-1}}{Z_{N}}\right )^{i}}. \label{34-000}
\end{eqnarray}
By using Eq. (\ref{34-00}), one can   write the sum in Eqs.
(\ref{19}) and (\ref{21.1}) for $\gamma e^{- \beta \epsilon_{p}} <
1$ as follows:
\begin{eqnarray}
\langle a^{+}_{p_{1}}a_{p_{2}}\rangle_{N,V \rightarrow \infty}  =
\frac{\delta_{p_1p_2}} { e^{\beta \epsilon_{p_{2}}} \gamma^{-1} -1}
\label{34}
\end{eqnarray}
and
\begin{eqnarray}
\langle  a^{+}_{p_{1}}a^{+}_{p_{2}}a_{p_{1}}a_{p_{2}}\rangle_{N,V
\rightarrow \infty}  = \frac{\delta_{p_1p_2}\delta_{p_2p_1} +
\delta_{p_2p_2}\delta_{p_1p_1}}{\left ( e^{\beta \epsilon_{p_{1}}}
\gamma^{-1} -1 \right) \left ( e^{\beta \epsilon_{p_{2}}}
\gamma^{-1} -1 \right)}.\label{35}
\end{eqnarray}
Note  that the above expressions  satisfy the principle of
thermodynamic equivalence between the canonical ensemble  and the
grand-canonical ensemble with  $e^{\beta \mu}=\frac{Z_{N-1}}{Z_{N}}$
(see also Ref. \cite{Pet}).

 Finally, redefining in the thermodynamic limit $\delta
(\textbf{k}-\textbf{k}')=(2 \pi)^{-3}V\delta_{kk'}$ and
$a(\textbf{k})=((2 \pi)^{-3}V)^{1/2}(2E(k))^{1/2}a_{k}$,
$E(k)=\sqrt{k^{2}+m^{2}}$, we get
\begin{eqnarray}
\langle a^{+}(\textbf{p}_{1})a(\textbf{p}_{2})\rangle  =
\frac{2E(p_{2})\delta(\textbf{p}_{1}-\textbf{p}_{2})} { e^{\beta
E(p_{2})} \gamma^{-1} -1} \label{36}
\end{eqnarray}
and
\begin{eqnarray}
\langle a^{+}(\textbf{p}_{1})a^{+}(\textbf{p}_{2})a(\textbf{p}_{1})a
(\textbf{p}_{2})\rangle  =
\frac{4E(p_{2})E(p_{1})((\delta(\textbf{p}_{1}-\textbf{p}_{2}))^{2}
+ (\delta(\textbf{0}))^{2})}{\left ( e^{\beta E(p_{1})} \gamma^{-1}
-1 \right) \left ( e^{\beta E(p_{2})} \gamma^{-1} -1 \right)}.
\label{37}
\end{eqnarray}
It is easily seen from Eqs. (\ref{36}) and (\ref{37})  that $\langle
a^{+}(\textbf{p}_{1})a^{+}(\textbf{p}_{2})a(\textbf{p}_{1})a
(\textbf{p}_{2})\rangle $ can be written as
\begin{eqnarray}
\langle a^{+}(\textbf{p}_{1})a^{+}(\textbf{p}_{2})a(\textbf{p}_{1})a
(\textbf{p}_{2})\rangle  = \nonumber \\ \langle
a^{+}(\textbf{p}_{2})a (\textbf{p}_{1})\rangle \langle
a^{+}(\textbf{p}_{1})a(\textbf{p}_{2})\rangle + \langle
a^{+}(\textbf{p}_{1})a(\textbf{p}_{1})\rangle \langle
a^{+}(\textbf{p}_{2})a(\textbf{p}_{2})\rangle. \label{35.1}
\end{eqnarray}

Equation (\ref{35.1}) is the particular case of the thermal Wick's
theorem \cite{Wick-1}. Evidently,  this is the consequence of
general results about the equivalency of canonical and
grand-canonical ensembles in the thermodynamic limit. However,  when
particle production occurs from extremely small systems with a
typical size of  $10^{-14}$ m created in relativistic nucleus and
particle collisions, one should use an approximation of the
thermodynamic limit with great caution especially in  correlation
femtoscopy analysis of the system's effective size.

\section{Two-boson correlations  at fixed multiplicities and finite volumes}

In the previous section we calculated the expectation values of
creation and annihilation operators  both  in a finite volume with
periodical boundary conditions and in the thermodynamic limit. We
demonstrated that only for the latter case is the thermal Wick's
theorem  applied in the canonical ensemble of non-interacting
particles with the fixed particle number constraint. Our aim in this
section is to adjust these results  for correlation femtoscopy
analysis of systems created in high-energy particle and nucleus
collisions. To do this, one needs to take into account, first,  that
experimentally measured observables are one- and two- particle
momentum spectra in the continuous-mode momentum representation,
and, second, that finite sizes measured by this method do not allow
one to treat these systems in the thermodynamic limit. Then,  to be
able to consider  such systems in a simple
approximation,\footnote{An exact approach should be based on a local
equilibrium statistical operator for finite systems, similar to how
it was realized in Ref. \cite{Wick-2} for the particular case of a
grand-canonical ensemble  in a locally equilibrated longitudinally
expanding system. However, for a canonical ensemble with a fixed
particle number, this is an especially nontrivial problem.} let us
assume that the Compton wavelength and  the thermal one are much
less than the size of the system. This  allows one to use the
continuous-mode momentum representation as an approximation to the
discrete momentum representation in a finite volume without invoking
the thermodynamic limit. To evaluate quantities of interest, we use
as the starting point Eqs. (\ref{19}) and (\ref{21.1}), where we
substitute $(2 \pi)^{-3}V\delta_{kk'}\rightarrow \delta_{V}
(\textbf{k}-\textbf{k}')$ and $a_{k} \rightarrow a(\textbf{k})((2
\pi)^{-3}V)^{-1/2}(2E(k))^{-1/2}$.   Note that the appearance of
$\delta_{V} (\textbf{k}-\textbf{k}')$ instead of $(2
\pi)^{-3}V\delta_{kk'}$ (or $\delta (\textbf{k}-\textbf{k}')$) does
not mean  modification of the commutation relation of the creation
and annihilation operators but just some modification of the
averaging of their products for the case when the spatial particle
number density, which can be calculated from
 the one-particle Wigner function (see, e.g., Ref.
\cite{Groot}), comes to be  not constant in full space but related
to a finite system.  Also, it is natural to assume  that the
corresponding density distribution is not sharp but has a  smooth
cutoff. The important example is that in a grand-canonical system
with a Gaussian spatial particle number density distribution one
should put
\begin{equation}
\delta_{V} (\textbf{k}-\textbf{k}') =
\frac{R^{3}}{(2\pi)^{3/2}}e^{-(\textbf{k}-\textbf{k}')^{2}R^{2}/2}
\label{delta}
\end{equation}
in Eqs. (\ref{12}) (\ref{15}) to reproduce the corresponding density
behavior and the Gaussian distribution of particle momentum
difference observed in Bose-Einstein two-particle  correlation
data.\footnote{The corresponding results for a specific case that
includes the symmetrization procedure for an initially
nonsymmetrized amplitude of boson radiation from factorized
independent (noncoherent) sources are presented in Ref.
\cite{Ledn}.}  Let us assume that the same substitution
(\ref{delta}) can be applied also in a canonical ensemble with fixed
particle multiplicity, Eqs. (\ref{19}) and  (\ref{21.1}).   Then the
parameter $R$ is related to the volume $V$ used in the discrete-mode
``box'' representation as follows:
\begin{eqnarray}
R^{3}=V(2\pi)^{-3/2}. \label{38}
\end{eqnarray}

The computation of  $\langle
a^{+}(\textbf{k})a^{+}(\textbf{k}')a(\textbf{k})a(\textbf{k}')\rangle_{N}$
and $\langle a^{+}(\textbf{k})a(\textbf{k})\rangle_{N}$  makes it
possible to obtain  the two-particle correlation function, which is
defined  as
\begin{eqnarray}
C_{N}(\textbf{p},\textbf{q})=C_{N} \frac{\langle
a^{+}(\textbf{p}_{1})a^{+}(\textbf{p}_{2})a(\textbf{p}_{1})a(\textbf{p}_{2})\rangle_{N}}{\langle
a^{+}(\textbf{p}_{1})a(\textbf{p}_{1})\rangle_{N}\langle
a^{+}(\textbf{p}_{2})a(\textbf{p}_{2})\rangle_{N}}, \label{39}
\end{eqnarray}
where ${\bf p}=({\bf p}_{1}+{\bf p}_{2})/2$, ${\bf q}={\bf
p}_{2}-{\bf p}_{1}$, and $C_{N}$ is the normalization constant; the
latter  is needed  to normalize the theoretical correlation function
in accordance with normalization that is applied by
experimentalists:  $C_{N}(\textbf{p},\textbf{q}) \rightarrow
 1$ for $|\textbf{q}| \rightarrow \infty$.
Substitution  of Eqs. (\ref{19}) and (\ref{21.1}) into Eq.
(\ref{39}) yields
\begin{eqnarray}
C_{N}(\textbf{p},\textbf{q})= C_{N}  \frac{(1 +
e^{-\textbf{q}^{2}R^{2}})\sum_{i=1}^{N-1} \sum_{j=1}^{N-i}
e^{-i\beta E (p_{2})-j \beta
E(p_{1})}\frac{Z_{N-i-j}}{Z_{N}}}{\sum_{i=1}^{N}\sum_{j=1}^{N}
e^{-i\beta E(p_{2})}e^{-j\beta
E(p_{1})}\frac{Z_{N-i}}{Z_{N}}\frac{Z_{N-j}}{Z_{N}}},  \label{40}
\end{eqnarray}
where we performed the transition to the continuous-mode
representation as is described above. Despite  its complexity, this
expression allows a trivial evaluation of the $|\textbf{q}|
\rightarrow \infty$ limit. Namely, because $E(p_{1})$ and $E(p_{2})$
tend to $\infty$ when $|\textbf{q}| \rightarrow \infty$ at fixed
$\textbf{p}$, we see that $C_{N}(\textbf{p},\textbf{q})\rightarrow
C_{N}\frac{Z_{N-2}Z_{N}}{Z_{N-1}^2}$ in this limit. If we now demand
that $C_{N}(\textbf{p},\textbf{q})\rightarrow 1$ when  $|\textbf{q}|
\rightarrow \infty$, we see that the normalization factor introduced
in Eq. (\ref{39}) is
\begin{eqnarray}
 C_{N} = \frac{Z_{N-1}^2}{Z_{N-2}Z_{N}}.
\label{41}
\end{eqnarray}

\begin{figure}[H]
     \centering
     \includegraphics[width=0.8\textwidth]{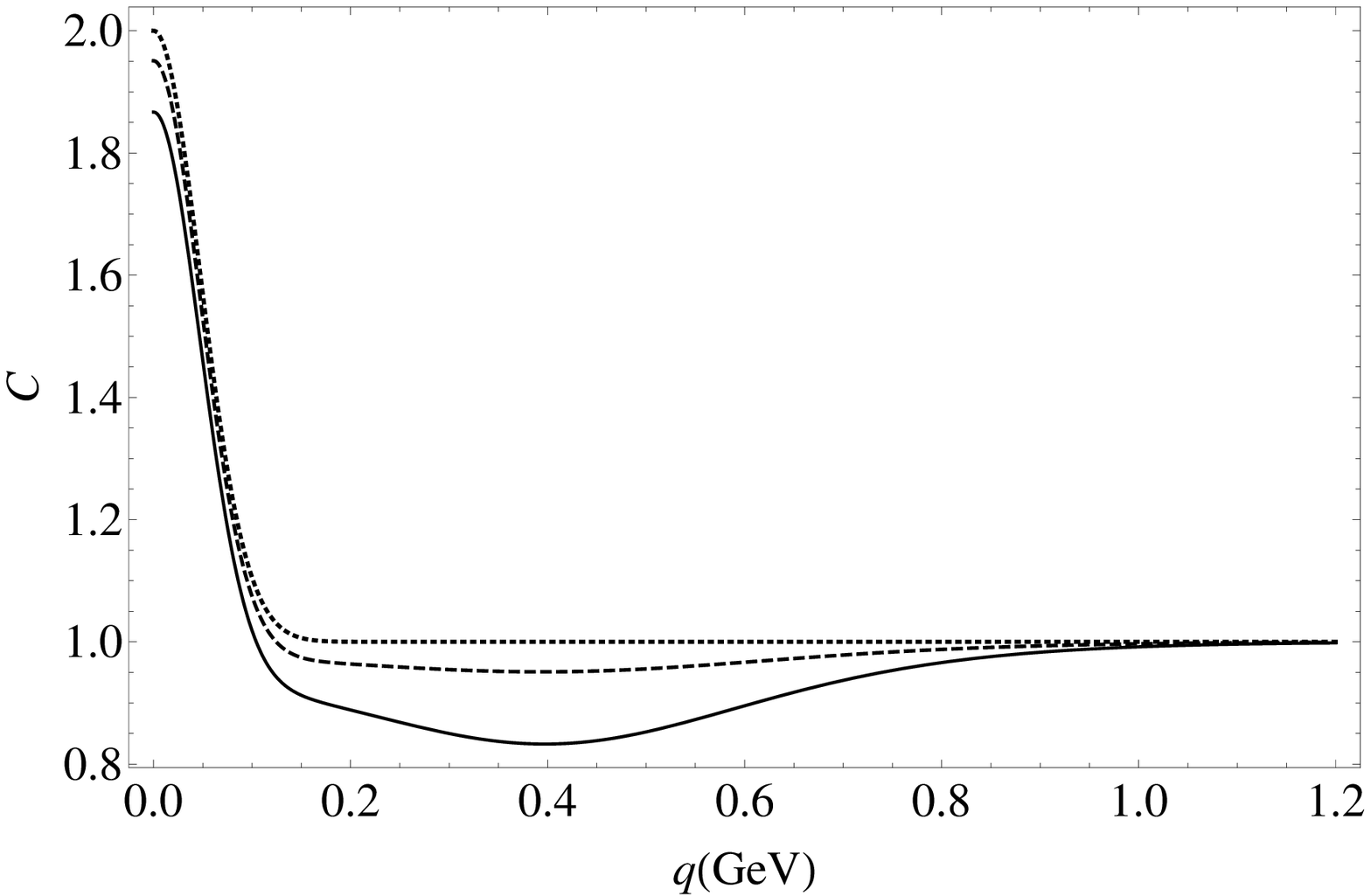}
     \caption{\small
The two-pion correlation functions $C(q_x=q,q_y=0,q_z=0;
p_x=p,p_y=0,p_z=0)$, with $p=0.2$ GeV/c, $R=3$ fm, and $T= 0.06$ GeV
in the case of the two-particle system: $N=2$. The dotted line
corresponds to the ``standard'' expression for the  pure
Bose-Einstein correlation function (CF) of the Gaussian source. The
dashed line is related to the CF when one-boson spectra in the
two-boson system is calculated from the two-particle spectra by
integrating it  over one of the momenta. The solid line corresponds
to our approximation based on Eq. (\ref{40}). } \label{fig:C-two}
\end{figure}

Note that systems with small particle numbers can be easily
investigated by exploiting recurrence relations for the partition
function and performing straightforward exact calculations.
 In Fig. \ref{fig:C-two} we illustrate and compare
the results for two-pion correlation functions $C(q_x=q,q_y=0,q_z=0;
p_x=p,p_y=0,p_z=0)$ (\ref{39}), with $p=0.2$ GeV/c, $R=3$ fm, and
$T= 0.06$ GeV in the case when the two-pion spectrum, see the
numerator of Eq. (\ref{40}), is calculated for the two-particle
system: $N=2$. Our aim here is, in particular, to compare the
results of the calculations of the two-particle momentum
correlations (\ref{39}) by means of Eq. (\ref{40}) with some other
prescriptions. First, we present the ``standard'' correlation
function $C(\textbf{p},\textbf{q})=1+\exp{(-\textbf{q}^2R^2)}$ that,
in fact, corresponds to the correlation function (\ref{40}) with
$N=2$ in the numerator and $N=1$ in the denominator (the prefraction
normalization constant $C_2$ is modified correspondingly). Such a
formal case was considered  in Ref. \cite{Sin-1}, and it was shown
that the above ``standard'' expression can be  obtained   only for a
fairly large system when one can ignore the quantum uncertainty
principle in particle radiation. It is worth   noting  that the same
``standard'' behavior of the two-particle correlation function takes
place in the grand-canonical ensemble for bosons \cite{Sin} when the
(effective) system sizes are much larger then the thermal de Broglie
wavelength \cite{Wick-2,Sin-1}.

The second result that we present for the two-particle correlation
function corresponds to the case when the one-particle pion spectra
($\langle a^{+}(\textbf{p}_{1})a(\textbf{p}_{1})\rangle_{2}$ and
$\langle a^{+}(\textbf{p}_{2})a(\textbf{p}_{2})\rangle_{2}$ in Eq.
(\ref{39})) are calculated  by integration of the two-particle
momentum spectra (numerator of Eq. (\ref{40})) over one of the
particle momenta.

The third result is based on our approximation (\ref{40})  of  the
correlation function of a finite system  without any modification.
One can see that the last result qualitatively reproduces the
previous one where the one-particle  spectra are calculated from the
two-particle ones in a direct way (that is, in fact, an exact
self-consistent approach) and the two-particle spectra are taken in
our finite-system approximation (numerator of Eq. (\ref{40})).
Namely, in both approaches the correlation functions are suppressed
(their intercepts are reduced), they lose the  Gaussian form even
for the Gaussian source, and the correlation function
$C(\textbf{p},\textbf{q})$ becomes less than unity at intermediate
$q$ values and approaches the limiting value of $1$ from below. Such
a similarity of the results supports our approximation (\ref{40}) of
the two-particle momentum  distribution function.\footnote{It is
worth  noting  that the same peculiarities of the two-particle
momentum correlations for fixed-$N$ systems   were obtained in Ref.
\cite{Ledn}. The ultimate reason for such behavior is  the violation
of the Wick's theorem for heat systems with fixed multiplicities. }

To see the corresponding  effects  in the systems with finite but
rather large $N$, as  can happen in relativistic particle and
nucleus collisions, let us first write the recurrence relation
(\ref{25}) as
\begin{eqnarray}
Z_{N} = \frac{1}{N}\sum_{i=0}^{N-2}Z_{i} Z_{1}((N-i)\beta)+
\frac{1}{N}Z_{N-1}Z_{1}(\beta). \label{26}
\end{eqnarray}
Utilizing again Eq. (\ref{25}), one can express $Z_{N-1}$ as
\begin{eqnarray}
Z_{N-1} = \frac{1}{N-1}\sum_{i=0}^{N-2}Z_{i} Z_{1}((N-i-1)\beta) .
\label{27}
\end{eqnarray}
Thus we observe that
\begin{eqnarray}
\frac{Z_{N}}{Z_{N-1}} = \frac{N-1}{N}\frac{\sum_{i=0}^{N-2}Z_{i}
Z_{1}((N-i)\beta)}{\sum_{i=0}^{N-2}Z_{i}
Z_{1}((N-i-1)\beta)}+\frac{1}{N}Z_{1}(\beta). \label{28}
\end{eqnarray}
Now, using Eq. (\ref{25.1}), one can change the order of summations
in Eq. (\ref{28}) and get
\begin{eqnarray}
\frac{Z_{N}}{Z_{N-1}} = \alpha_{N-2} \exp{(-\beta m) } \frac{N-1}{N}
+ \frac{Z_{1}(\beta)}{N}, \label{31}
\end{eqnarray}
where
\begin{eqnarray}
\alpha_{N-2}=\frac{\sum_{p}f_{N-2}(\beta \epsilon_{p})\exp{(-\beta
\epsilon_{p}+\beta m)}}{\sum_{p}f_{N-2}(\beta
\epsilon_{p})},\label{29}\\
f_{N-2}(\beta
\epsilon_{p})=\sum_{i=0}^{N-2}Z_{i}(\beta)\exp{(-(N-i-1)\beta
\epsilon_{p})}, \label{30}
\end{eqnarray}
and $m$ is a particle  mass. It is  apparent from Eq. (\ref{29})
that $\alpha_{N-2} < 1$.
 Now, note that in the continuous-mode representation $\sum_{p}\rightarrow
\frac{V}{(2\pi)^{3}}\int d^{3}p $, $\epsilon_{p}\rightarrow
E(p)=\sqrt{p^{2}+m^{2}}$, and Eq. (\ref{31}) takes  the following
form,
\begin{eqnarray}
\frac{Z_{N}}{Z_{N-1}} =
\frac{1}{(2\pi)^{3}}\frac{V}{N}I(\beta)+\alpha_{N-2} \exp{(-\beta m)
}, \label{42}
\end{eqnarray}
where
\begin{eqnarray}
I(\beta) \equiv \int d^{3}p \exp{(- \beta E(p))}=4\pi \beta^{-1}
m^{2} K_{2}(\beta m)  \label{33}
\end{eqnarray}
and $V=(2\pi)^{3/2}R^{3}$, see Eq. (\ref{38}). Taking into account
that $\alpha_{N-2}<1$ and assuming the low particle number density
approximation,
\begin{eqnarray}
\frac{1}{(2\pi)^{3}}\frac{V}{N}I(\beta)\gg 1, \label{33-1}
\end{eqnarray}
we then get\footnote{An attentive reader will notice that Eq.
(\ref{43}) is associated with the Boltzmann approximation in the
thermodynamic limit.}
\begin{eqnarray}
\frac{Z_{N-1}}{Z_{N}} \simeq (2\pi)^{3}\frac{N}{V}I^{-1}(\beta) \ll
1. \label{43}
\end{eqnarray}
It is easily seen that $\frac{Z_{N-i-1}}{Z_{N}}=
\frac{Z_{N-i-1}}{Z_{N-i}}...\frac{Z_{N-1}}{Z_{N}}\ll
\frac{Z_{N-1}}{Z_{N}}\ll 1$, giving from Eqs. (\ref{40}) and
(\ref{41}) the following approximate formula,
\begin{eqnarray}
C(\textbf{p},\textbf{q}) \simeq \frac{1+ (e^{-\beta E
(p_{1})}+e^{-\beta E (p_{2})})\frac{Z_{N-3}}{Z_{N-2}}}{1+(e^{-\beta
E(p_{1})}+e^{-\beta E(p_{2})})\frac{Z_{N-2}}{Z_{N-1}}} \left (1 +
e^{-\textbf{q}^{2}R^{2}} \right ), \label{44}
\end{eqnarray}
which can be further simplified as
\begin{eqnarray}
C(\textbf{p},\textbf{q}) \simeq \left(1+ (e^{-\beta E
(p_{1})}+e^{-\beta E (p_{2})})\left(\frac{Z_{N-3}}{Z_{N-2}} -
\frac{Z_{N-2}}{Z_{N-1}}\right ) \right) \left (1 +
e^{-\textbf{q}^{2}R^{2}} \right ). \label{45}
\end{eqnarray}
Taking into account Eq. (\ref{43}), the above expression reads
\begin{eqnarray}
C(\textbf{p},\textbf{q}) \simeq \left(1-\frac{(2\pi)^{3}}{VI(\beta)}
(e^{-\beta E (p_{1})}+e^{-\beta E (p_{2})})\right)\left(1 +
e^{-\textbf{q}^{2}R^{2}}\right). \label{46}
\end{eqnarray}
To see qualitative peculiarities of the above expression explicitly,
let us rewrite  Eq. (\ref{46}) for $\beta m \gg 1$.\footnote{Our
conclusions are valid, in fact,  for any $\beta m$ values, because
only the quantitative strength of the effect depends on the  $\beta
m$ value.} Then $I(\beta)\approx \lambda_{T}^{-3} e^{- \beta m}$,
where $\lambda_{T}=(2 \pi m T)^{-1/2}$ is the so-called thermal
wavelength, $T=1/\beta$, and Eq. (\ref{46}) can be simplified as
\begin{eqnarray}
C_{N}(\textbf{p},\textbf{q}) \simeq
(1-(2\pi)^{3}\frac{\lambda_{T}^3}{V}e^{\beta m} (e^{-\beta E
(p_{1})}+e^{-\beta E (p_{2})}))(1 + e^{-\textbf{q}^{2}R^{2}}).
\label{47}
\end{eqnarray}
It is instructive to compare the above expression  with a typical
experimental parametrization of the two-boson Bose-Einstein
correlation function,  which looks like
\begin{eqnarray}
 C_{N}(\textbf{p},\textbf{q})=1+\lambda(p) \exp{(-R_{G}^{2}\textbf{q}^{2})}. \label{47.1}
\end{eqnarray}
Here  $0 \leq\lambda (p)\leq 1$   describes the correlation
strength, $R_{G}$ is the Gaussian interferometry radius, and to
allow comparison with our simple model we simplify the real
experimental three-dimensional  parametrization by assuming
spherical symmetry in Eq. (\ref{47.1}). To do this comparison, first
note that $\lambda (p)= C_{N}(\textbf{p},\textbf{0}) - 1$, and one
can see from Eq. (\ref{47}) that the intercept
$C_{N}(\textbf{p},\textbf{0})$  is
\begin{eqnarray}
C_{N}(\textbf{p},\textbf{0}) \simeq
2(1-2(2\pi)^{3}\frac{\lambda_{T}^3}{V} e^{-\beta E (p)+\beta m}).
\label{48}
\end{eqnarray}
Therefore for the correlation strength parameter  $\lambda (p)$ we
get in the limit of small particle number densities (\ref{33-1})
\begin{eqnarray}
\lambda (p) \simeq 1- 4 e^{-\beta E (p)+ \beta
m}(2\pi)^{3}\frac{\lambda_{T}^3}{V}. \label{49}
\end{eqnarray}
Note that the Bose-Einstein correlations are suppressed, $\lambda
(p)<1$, and that $\lambda (p)\rightarrow 1$   when $V \rightarrow
\infty$; therefore Eq. (\ref{49}) represents the finite-volume
effect in a canonical ensemble with fixed particle multiplicity. One
can see that the suppression becomes stronger when the system size
tends to the thermal de Broglie wavelength. It has been  shown in
the model of independent incoherent emitters that $\lambda
\rightarrow 0$ in a {\it finite} system at fixed multiplicity, if
particle number tends to infinity \cite{Ledn, Pratt-laser}. It is
interesting to note that $\lambda$  is usually interpreted as a
measure of coherence  in a theoretical model: $\lambda=0$ for a pure
quantum state and $0<\lambda\leq 1$  for a mixed quantum state (see,
e.g., Refs. \cite{HBT,Sin-1}). However, in the case of a thermal
system with fixed multiplicity, the suppression of the correlation
function is the result of violation of the Wick's theorem. Then the
coherence effects need  an additional specific treatment
\cite{Sin-1}.

Using Eq. (\ref{49}), we see that Eq. (\ref{47}) can be written in
the form
\begin{eqnarray}
C(\textbf{p},\textbf{q})\simeq \frac{1}{2}(1+\lambda
(p)f(p,q))(1+e^{-R^{2}\textbf{q}^{2}}) \label{50},
\end{eqnarray}
where the factor
\begin{eqnarray}
f(p,q)=\frac{ e^{-\beta E (p_{1})}+ e^{-\beta E (p_{2})}}
{2e^{-\beta E (p)}} \label{51}
\end{eqnarray}
results in non-Gaussian behavior of the correlation function  with
respect to $|\textbf{q}|$. Namely, for $|\textbf{p}| \gg
|\textbf{q}|$ and $|\textbf{p}| \gg m$,
\begin{eqnarray}
\frac{ e^{-\beta E (p_{1})}+ e^{-\beta E (p_{2}}} {2e^{-\beta E
(p)}}\approx \exp {\left (-\frac{q^2}{8T\sqrt{p^{2}+m^{2}}}\right )
}\cosh {\left (\frac{pq}{2T\sqrt{p^{2}+m^{2}}}\right )} .
\label{51.1}
\end{eqnarray}

Equations (\ref{47.1}), (\ref{50}), and (\ref{51}) allow us to
relate $R$ and $R_{G}$ for  small $|\textbf{q}|$, when
$R^{2}\textbf{q}^{2}\ll 1$ and $R_{G}^{2}\textbf{q}^{2}\ll 1$. Then,
expanding $C(\textbf{p},\textbf{q})$ in $\textbf{q}$ and keeping the
first term in the Taylor series, we get from Eq. (\ref{50})
\begin{eqnarray}
C(\textbf{p},\textbf{q})\simeq \frac{1}{2}(1+ \lambda
(p))(2-R^{2}\textbf{q}^{2}), \label{52}
\end{eqnarray}
and we also get from Eq. (\ref{47.1})
\begin{eqnarray}
C(\textbf{p},\textbf{q})\simeq 1+ \lambda
(p)(1-R_{G}^{2}\textbf{q}^{2}). \label{53}
\end{eqnarray}
Here we assume that the  system size is large enough in comparison
with the thermal wavelength, etc., and replace $f(p,q)$ by $1$.
After equating Eqs. (\ref{52}) and (\ref{53}) we get eventually
\begin{eqnarray}
R_{G}^{2}=\frac{1+\lambda(p)}{2\lambda(p)} R^{2}. \label{54}
\end{eqnarray}
We remind the reader that this rough approximation is valid only
when $\lambda$ is close to unity. It is interesting to note that the
Gaussian interferometry  radius exhibits a decrease  as  the pair
momentum increases that is similar to the well-known behavior of the
homogeneity length that is associated with the Gaussian
interferometry radius in a locally equilibrated expanding system
\cite{Sin-2}.

\section{Conclusions}

In  this paper,  we have analyzed the single- and two- particle
momentum spectra for finite canonical systems of noninteracting
particles with a fixed particle number constraint. We find that the
corresponding expressions satisfy the  thermal Wick's theorem in the
thermodynamic limit  and do not satisfy it in the general finite
case. Our analysis implies that contrary to traditional beliefs (see
e.g. Ref. \cite{Heinz}), decomposition of two-particle distribution
functions through one-particle ones  is questionable for the  class
of events with fixed multiplicities originating from small thermal
systems.

Furthermore,   we evaluated the two-particle correlation function in
a low particle number density approximation in a canonical ensemble
model. An analysis of the two-particle correlation function
indicates that even for Gaussian sources the correlation function is
non-Gaussian and reaches values less than unity in some intermediate
region of relative momentum of particles $q$;  the apparent source
size (interferometry radius extracted from the typical experimental
parametrization for small values of $|\textbf{q}|$) decreases with
the half-sum pair momentum $|\textbf{p}|$ and at large
$|\textbf{p}|$ reaches the static Gaussian source value.

It was found that the finite-size effect in thermal systems with
fixed multiplicity  results in a reduction  of the intercept of the
correlation function. In approximation of small particle number
density,  the ``coherence parameter'' $\lambda$ becomes smaller than
unity and decreases when the system size approaches the thermal de
Broglie wavelength.  As $|\textbf{p}|$ increases the parameter rises
gradually and reaches the constant value of $1$.  It is interesting
to note that  measured  with various methods of analysis by
different LHC collaborations \cite{HBT-pp} values of $\lambda$ in
$p+p$ collisions are smaller than unity and exhibit a decrease as
the average momentum of the pair increases. This is at variance with
the behavior of $\lambda$ in a canonical ensemble model with a fixed
particle number constraint and could  be caused by a specific
nonthermal mechanism of particle production in $p+p$ collisions,
e.g., by pions originated from resonances, but analysis of the
latter goes beyond the scope of the present article.

In the present work, to make the problem tractable, we used a simple
static canonical ensemble model, while the particle-emitting sources
produced in high-energy nucleus and particle collisions are
expanding, interactions between the emitted particles  are rather
complicated and include the resonance decays, etc. Therefore,
further investigations based on a more realistic model of an
evolving source could  be of great interest. However, we hope that
the  present analysis sheds light on the additional causes  (besides
the purity, long-lived resonances and coherence) of the suppression
of the correlation function  intercept value  expressed by the
parameter $\lambda$. Namely, we found that this parameter is a
measure of the degree of the Wick's theorem violation in a thermal
system with a fixed particle number constraint. The results of our
analysis may be useful for interpretation of the results of the
correlation femtoscopy  of events with fixed multiplicities and also
for event-generator modeling of the two-particle Bose-Einstein
correlations arising in small thermal systems created in
relativistic nucleus and particle collisions.

\begin{acknowledgments}
Yu.S. is grateful to ExtreMe Matter Institute EMMI/GSI for support
and a visiting professor position. This research was carried out
within the scope of the EUREA: European Ultra Relativistic Energies
Agreement (European Research Group: ``Heavy Ions at
Ultrarelativistic Energies'') and is supported by the National
Academy of Sciences of Ukraine, Contract No. F7-2016.
\end{acknowledgments}

\begin{thebibliography}{99}

\bibitem{KopPodg} M.I. Podgoretsky, Fiz. Elem. Chast. At. Yad. \textbf{20},
628 (1989) [Sov. J. Part. Nucl. \textbf{20}, 266 (1989)].

\bibitem{HBT} U.A. Wiedemann, U. Heinz,  Phys. Rep. \textbf{319}, 145 (1999); R.M.
Weiner, Phys. Rep. \textbf{327},  249   (2000); \textit{Introduction
to Bose-Einstein Correlations and Subatomic Interferometry} (Wiley,
New York, 2000); M. Lisa, S. Pratt, R. Soltz,  U. Wiedemann, Annu.
Rev. Nucl. Part. Sci. \textbf{55}, 357 (2005).

\bibitem{Sin-1} Yu.M. Sinyukov, V.M. Shapoval, Phys. Rev. D \textbf{87}, 094024 (2013).

\bibitem{Heinz} C. Plumberg, U. Heinz, Phys. Rev. C \textbf{91}, 054905 (2015) [arXiv:1503.05605];
 Phys. Rev. C \textbf{92}, 044906 (2015)
 [arXiv:1507.04968];  arXiv:1512.07631
[nucl-th].

\bibitem{HBT-pp} K. Aamodt \textit{et al} (ALICE Collaboration), Phys. Rev. D \textbf{84}, 112004
(2011); S.S. Padula (for the CMS Collaboration), arXiv:1502.05757
[nucl-ex]; ATLAS Collaboration, Eur. Phys. J. C \textbf{75}, 466
(2015) [arXiv:1502.07947].

\bibitem{Hydro} U. Heinz, J. Phys.: Conf. Ser. \textbf{455},
012044 (2013)   [arXiv:1304.3634]; U. Heinz, R. Snellings, Annu.
Rev. Nucl. Part. Sci. \textbf{63}, 123 (2013) [arXiv:1301.2826]; C.
Gale, S. Jeon, B. Schenke, Int. J. Mod. Phys. A  \textbf{28},
1340011  (2013) [arXiv:1301.5893]; P. Huovinen,   Int. J.  Mod.
Phys. E \textbf{22}, 1330029 (2013) [arXiv:1311.1849]; R. Derradi de
Souza, T. Koide, T. Kodama,  Prog. Part. Nucl. Phys. \textbf{86}, 35
(2016) [arXiv:1506.03863].

\bibitem{urqmd} S.A. Bass \textit{et al}., Prog. Part. Nucl. Phys. \textbf{41}, 255
(1998); M. Bleicher \textit{et al.}, J. Phys. G \textbf{25}, 1859
(1999).

\bibitem{flow-pp} K. Werner,  Iu. Karpenko, T.
Pierog, M. Bleicher, K. Mikhailov,   Phys. Rev. C \textbf{83},
044915 (2011) [arXiv:1010.0400]; V.M. Shapoval, P. Braun-Munzinger,
Iu.A. Karpenko, Yu.M. Sinyukov, Phys. Lett. B \textbf{725}, 139
(2013)[arXiv:1304.3815]; T. Kalaydzhyan, E. Shuryak, Phys. Rev. C
\textbf{91}, 054913 (2015) [arXiv:1503.05213]; Y. Hirono, E.
Shuryak, Phys. Rev. C \textbf{91}, 054915 (2015) [arXiv:1412.0063].

\bibitem{CFp} F. Cooper, G. Frye, Phys. Rev. D \textbf{10}, 186
(1974).

\bibitem{Sin} Yu.M. Sinyukov, B. Lorstad, Z. Phys. C \textbf{61}, 587 (1994).

\bibitem{Ledn} R. Lednicky, V. Lyuboshitz, K. Mikhailov, Yu. Sinyukov, A. Stavinsky, and B. Erazmus,
Phys. Rev. C  \textbf{61}, 034901 (2000) [arXiv:nucl-th/9911055].

\bibitem{Wick-1} C. Bloch, C. De Dominicis, Nucl. Phys. \textbf{7}, 459
(1958); M. Gaudin, Nucl. Phys. \textbf{15}, 89 (1960).

\bibitem{Bog} N.N. Bogolubov, N.N. Bogolubov, Jr., \textit{An Introduction to
 Quantum Statistical Mechanics} (Gordon and Breach, New York, 1992).

\bibitem{Groot}   S.R. de Groot, W.A. van Leeuwen, Ch. G. van Weert,
\textit{Relativistic Kinetic Theory} (North-Holland, Amsterdam,
1980).

\bibitem{Wick-2} Yu.M. Sinyukov, Preprint ITP-93-8E (unpublished); Heavy Ion Phys. \textbf{10}, 113 (1999)
[arXiv:nucl-th/9909018].

\bibitem{Recurr-2} W.J. Mullin, J.P. Fern\'{a}ndez, Am. J. Phys.  \textbf{71},  661 (2003)
[arXiv:cond-mat/0211115].

\bibitem{Recurr-3} W. Magnus, F. Brosens, arXiv:1505.04923
[cond-mat.stat-mech].

\bibitem{Recurr-1} P.T. Landsberg, \textit{Thermodynamics} (Interscience, New York, 1961);
P. Borrmann and G. Franke, J. Chem. Phys \textbf{98}, 2484 (1993).

\bibitem{Pet} D.Y. Petrina,\textit{ Mathematical Foundations of Quantum Statistical
Mechanics: Continuous Systems} (Springer, Dordrecht,  1995).

\bibitem{Pratt-laser} S. Pratt, Phys. Lett. B \textbf{301}, 159 (1993).

\bibitem{Sin-2} Yu.M. Sinyukov,  Nucl. Phys. A \textbf{566}, 589c (1994); Yu.M. Sinyukov,
in: \textit{Hot Hadronic Matter: Theory and Experiment} edited by J.
Letessier, H.H. Gutbrod, and J. Rafelski (Plenum, New York, 1995),
p. 309; S.V. Akkelin, Yu.M. Sinyukov,  Phys. Lett. B \textbf{356},
525 (1995); Z. Phys. C \textbf{72},  501 (1996).


\end{thebibliography}
\end{document}